\documentstyle[11pt,twoside,jltp,epsf]{article}

\title{Time-Dependent Transport in Mesoscopic Structures}

\author{Markus B\"uttiker\address{Department of Theoretical Physics, University of Geneva,
1211 Geneva, Switzerland}}

\runninghead{Markus B\"uttiker}{Time-Dependent Transport}
       
\begin{document}

\begin{abstract}
A discussion of recent work on time-dependent transport in mesoscopic 
structures is presented. The discussion emphasizes the 
use of time-dependent transport to gain information on the charge 
distribution and its collective dynamics. We discuss the RC-time 
of mesoscopic capacitors, the dynamic conductance of quantum point contacts
and dynamic weak localization effects in chaotic cavities. We review 
work on adiabatic quantum pumping and photon-assisted transport, and 
conclude with a list which demonstrates the wide range of problems
which are of interest.  

\end{abstract}

\maketitle


\section{Introduction}

Time-dependent transport in mesoscopic physics is a fascinating 
subject which yields a wealth of information which cannot be obtained 
otherwise. Electric fields, induced through oscillating voltages 
at the contacts or gates defining the sample, or induced 
by time-dependent magnetic fluxes,
couple to the charge distribution 
of the mesoscopic structure. Time-dependent transport 
is thus an investigation of the charge distribution and 
its dynamics. 
In addition to the purely scientific interest, the investigation 
of time-dependent transport in small structures is important if 
applications are envisioned, 
such as in quantum computing, or in more practical problems such as 
capacitance and current standards [\citen{flen}]. In all possible applications 
we are interested not in the stationary-time independent behavior 
of small structures, but in driving them from one state to another 
and in doing this as fast as possible. 

This work emphasizes a number of basic issues, provides some 
comments on the existing literature and attempts to make a few
suggestions for further research. The paper is not a scholarly review 
of the field. Nevertheless, due to the variety of topics addressed, 
we touch on a large portion of the recent literature and it is hoped 
that the article can thus also serve as a useful guide to the more recent 
work in this field. 

Time-dependent phenomena can be roughly classified 
depending on whether they are a consequence of {\it external forcing}
(externally applied voltages, time-dependent fluxes) 
or whether they occur {\it spontaneously} (thermal frequency 
frequency-dependent fluctuations, 
frequency-dependent shot noise). 
We can further distinguish, whether we deal with a linear process
and analyze {\it dynamic susceptibilities} (dynamic conductance, 
magnetic susceptibilities) or whether we deal with a {\it non-linear} 
process (rectification, photon-assisted tunneling). 
Furthermore it is useful to distinguish between low frequency 
phenomena where we follow {\it adiabatically} a sequence of
equilibrium states and a high-frequency {\it non-adiabatic} regime 
where the system is driven far from the ground state. 
These are already sixteen categories 
and it is clear that here we cannot 
address all of them. In addition we can combine 
different categories such as the investigation of shot noise in the 
presence of photon-assisted tunneling which is an example 
of an externally forced system in which we are in the non-linear,
non-adiabatic regime, but are interested in a spontaneous process 
(the fluctuation spectrum). It is clear that we cannot provide a 
reasonable discussion of all these diverse phenomena. 
To limit the scope, we consider only externally forced phenomena: 
spontaneous dynamic processes (fluctuations) in mesoscopic 
systems are reviewed in Ref. [\citen{blbu}]. 
 
\section{Basic considerations}

\subsection{Formulation of the problem} 

It is useful to consider first some basic aspects of the problem at hand. 
We ask: are there some general principles which we can or even must use when 
formulating a description of a time-dependent process? 
Already, when faced with the task to determine the equilibrium 
electrostatic potential of a mesoscopic structure, we become aware of the 
fact that we must look beyond the conductor whose state is of 
immediate interest to us. In a typical mesoscopic structure 
the equilibrium electrostatic potential depends not only on the conductor 
itself but also on the other nearby electric charges provided by 
donors or acceptors, by gates and by contacts.  To find the equilibrium 
electrostatic potential such additional nearby conductors must necessarily 
be part of the consideration. This fact is of particular importance 
in time-dependent transport, since what counts is not the externally applied 
field (presumed to be known) but the total electric field generated 
by all the relevant charges, whether they are within the conductor 
or away from it, on a gate or on the surface of a reservoir. 
Unlike in dc-transport where we get away with investigating particle 
currents, the total current ${\bf j}({\bf r})$ in time-dependent transport is 
the sum of the displacement current $(\epsilon_{L}/4\pi) 
\partial {\bf E}({\bf r})/\partial t$ and the particle current $j_{p} ({\bf r})$,
\begin{equation}
{\bf j}({\bf r}) = 
(\epsilon_{L}/4\pi) \partial {\bf E}({\bf r})/\partial t +  j_{p} ({\bf r}) .
\label{jtot}
\end{equation} 
Here $\epsilon_{L}$ is the dielectric constant (for simplicity taken 
to be space and time-independent). 
The total current is conserved, 
\begin{equation}
div {\bf j}({\bf r}) = 0. 
\label{divj}
\end{equation} 
Eqs. (\ref{jtot}) and (\ref{divj}) are a consequence of the 
continuity equation and the Poisson equation. Eq. (\ref{divj})
states that along a line that is tangential to the current vector ${\bf j}$, 
the length of this vector is an invariant. Like the conservation law 
of energy permits the transformation of kinetic energy into potential 
energy, so similarly here, we are permitted to transform 
particle current into displacement current and vice versa. 
Very importantly, while the particle current exists only 
inside electric conductors, the displacement current is not limited to the 
conductor, but exists wherever we have a time-dependent electric field. 

It is the total current which counts experimentally, not the 
particle current. This is particularly clear, if we can assume that 
all electromagnetic fields are localized. This assumption underlies 
the electrical engenieering networks composed of $R, C, L$ elements 
and possibly more complicated non-linear elements. As a consequence 
of the localization of the electromagnetic fields, the currents 
at the terminals of such a network add up to zero (there is overall
current conservation) and the sum of all charges in the network is 
also conserved. The localization of the electric field means that any 
field line which emanates from the conductor terminates a)
again on the conductor, b) at a nearby gate or capacitor which is 
included in our consideration or c) at a reservoir (electrical contact)
which must also be included into our consideration. The localization of 
electric fields means that we can find a volume,
denoted $V_{\Omega}$, large enough, such that the electric flux through the
surface of this 
Gauss volume vanishes. 
Naturally, this implies that the total charge $Q_{\Omega}$ within 
this Gauss volume vanishes and implies that the sum of all currents 
flowing in and out of this volume must add up to zero. It is 
thus reasonable to demand that we should provide a description 
of time-dependent transport such that the overall charge vanishes 
\begin{equation}
Q_{\Omega} (t) = 0
\label{qcons}
\end{equation} 
and such the sum of all currents at all 
the contacts (labeled $\alpha = 1, 2, 3, $) of the contacts of the 
conductor and nearby capacitors adds up to zero, 
\begin{equation}
\sum_{\alpha} I_{\alpha}(t)= 0 .
\label{curcons}
\end{equation} 
Only if these conditions are fulfilled do we get answers which 
depend only on potential differences (answers which are gauge invariant). 

In the simplest case the Gauss volume also coincides with the mesoscopic 
region in which phase-coherent electron motion is relevant. The Gauss volume 
separates then the mesoscopic region from the exterior macroscopic circuit
to which we can apply the usual engineering description in terms of 
$R$, $C$, $L$ elements, current and voltage sources, noise  etc. However, 
it should be noticed that this is not the only point of view: We might insist 
on treating the mesoscopic system and the external circuit on the same footing.
For instance in the circuit considered in Ref. [\citen{jack}] the potential 
distribution depends on the location of the battery vis-a-vis the 
conductor. On the quantum level, 
such an approach would demand that we write down  
Hamiltonians for current and voltage sources, microwave generators, etc.

\subsection{Mesoscopic capacitors}

\begin{figure}
%
\epsfxsize=7cm
\centerline{\epsffile{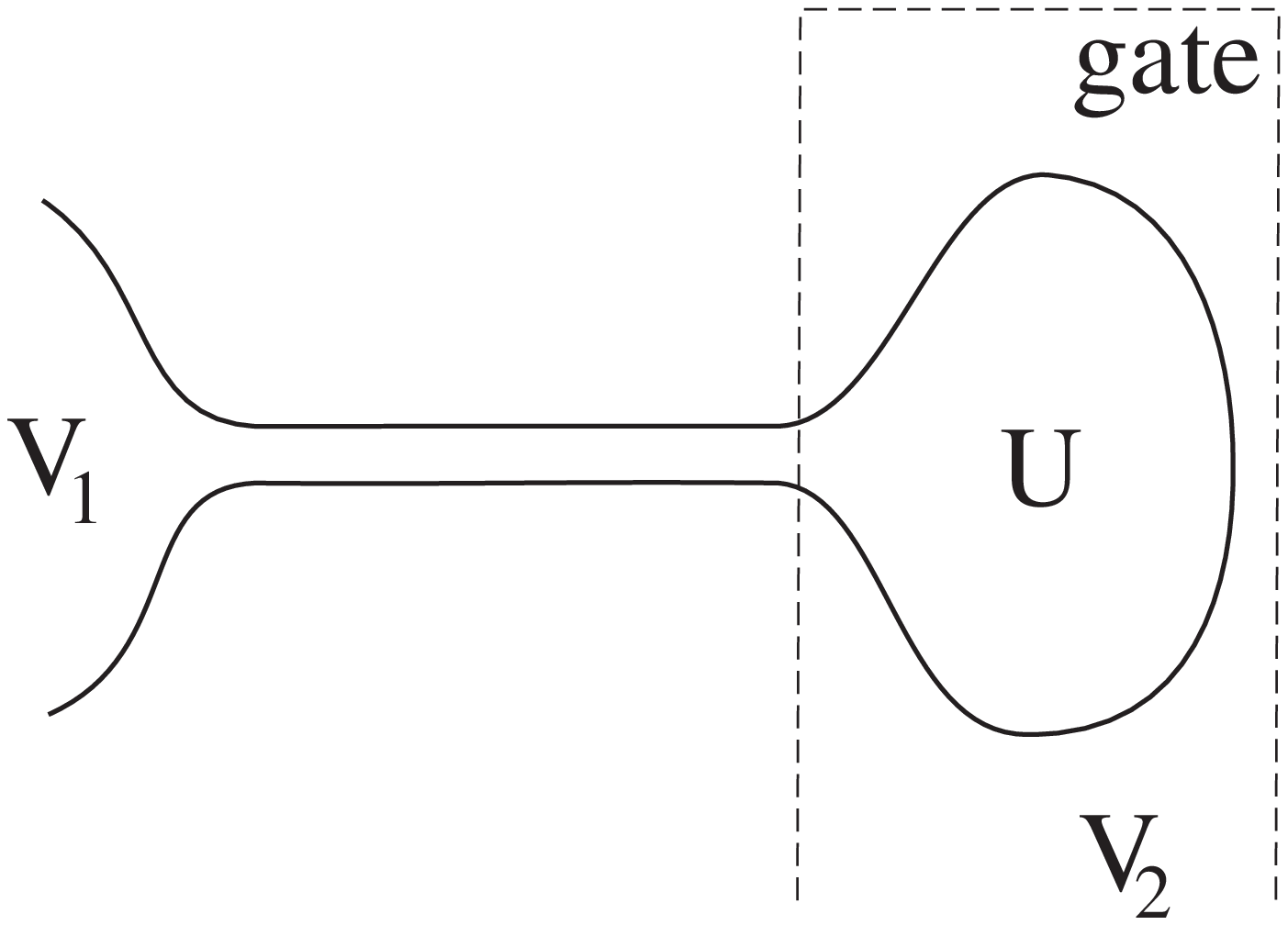}}
%
\caption{Mesoscopic capacitor connected via a single lead to an electron reservoir 
and capacitively coupled to a gate. $V_1$ and $V_2$ are the potentials applied
to the contacts, $U$ is the electrostatic potential of the cavity.  
}  
\label{mesocap}
\end{figure}

One of the elementary distinctions between dc-transport and ac-transport 
is that we can drive current not only through particle transport but also
through a displacement and thus can {\em correlate} particle 
currents in 
conductors which are not connected by a dc-conductance path. This aspect of 
dynamic conductance is perhaps most simply 
illustrated by considering a mesoscopic 
capacitor. Fig. \ref{mesocap} shows a mesoscopic cavity
connected only via one lead to an electron reservoir 
and separated from a back gate by an insulating layer. 
Note that there is no dc-transport possible in this structure.
We are interested in the dynamic conductance 
$G_{\alpha\beta}(\omega) = dI_{\alpha}(\omega)/
dV_{\beta} (\omega)$ which gives the current at contact 
$\alpha$ in response to an oscillating potential at contact $\beta$. 
We present the solution obtained in Ref. [{\citen{btp}}]. 
For simplicity, we assume that we can consider the electrostatic 
potential $U(\omega)$ in the cavity as uniform. Furthermore, we follow
the literature on the Coulomb blockade and instead of the Poisson equation
describe the relation between the charge on the cavity and the 
potential with the help of a geometrical capacitance $C$. 
The gate is described 
as macroscopic conductor. 
In response to an oscillating potential $dU(\omega)$ 
on the cavity and an oscillating gate voltage $dV_{2}(\omega)$
we have an oscillating charge on the cavity given 
by $dQ(\omega) = C(dU(\omega) - dV_{2}(\omega))$. 
We have to find $dU(\omega)$. 
Ref. \citen{btp} now considers first the response of non-interacting
carriers to an oscillation of the potential at contact $1$ assuming 
that the potential $U$ in the cavity is held fixed. This response is 
\begin{equation}
G^{0}(\omega) = \frac{e^{2}}{h} 
\int dE \, Tr [{\bf 1} - {\bf s}^{\dagger}(E) {\bf s}(E+\hbar \omega)]
\frac{f(E) -f(E +\hbar \omega)}{\hbar \omega}
\label{0con}
\end{equation}
where ${\bf s}$ is the scattering matrix which relates the incident 
current amplitudes in reservoir $1$ to the outgoing 
current amplitudes in reservoir $1$. ${\bf 1}$
is the unit matrix with dimension
equal to the number of scattering channels. 
The trace is just the sum over all
scattering channels. $f$ is the Fermi function in reservoir $1$.
Note that $G^{0}(0) = 0$ since the scattering matrix is unitary. 
The total current at contact $1$ is the result not only 
of the oscillating reservoir voltage $dV_{1}$ but also depends 
on the oscillating electric potential in the cavity. 
We write thus for the current at contact $1$,  
$dI_{1}(\omega) = 
G^{0}(\omega) dV_{1}(\omega)-i\omega \Pi (\omega) dU(\omega)$
with a response function $\Pi(\omega)$ which we now determine.
For the charge conserving answer which we seek, an overall potential 
shift cannot have an effect on the system. Thus if we subtract 
from all oscillating potentials $dU(\omega)$ we must find the same 
current as given above. This is the case only if $\Pi(\omega) = (-i/\omega)
G^{0}(\omega)$. Using this and observing that the current 
at contact $1$ is also the time-derivative of the charge on the cavity
gives
\begin{equation}
dI_{1}(\omega) = 
G^{0}(\omega) dV_{1}(\omega)- G^{0}(\omega) dU(\omega) 
= -i \omega C(dU(\omega) -dV_{2}(\omega)) .
\label{inv}
\end{equation}
This equation now determines the potential 
\begin{equation}
dU(\omega) = \frac{dV_{1} -i \omega C/G^{0}(\omega) dV_{2}}
{1 - i \omega C/G^{0}(\omega)} .
\label{pot}
\end{equation}
Inserting this potential back into Eq. (\ref{inv})
gives the conductance $G \equiv G_{11} = G_{22} = -G_{12} = -G_{21}$
of the interacting system, 
\begin{equation}
G(\omega) =  \frac{-i \omega C}
{1 - i \omega C/G^{0}(\omega)} .
\label{intg}
\end{equation}
We have now achieved a current conserving answer: 
Whether we measure at contact $1$ or $2$ we have to 
find the same current. 
We next would like to find the $RC$-time of the mesoscopic capacitor. 
To this end we consider $(1 - i \omega C/G^{0}(\omega))$ and expand it to 
first order in frequency. (This requires an expansion to second 
order in frequency of $G^{0}(\omega)$). 
This gives us a dynamic conductance of the form  
\begin{equation}
G(\omega) =  \frac{-i \omega C_{\mu}}
{1 - i \omega R_{q}C_{\mu}} .
\label{rcg}
\end{equation}
The dynamic conductance of the mesoscopic capacitor is, like that 
of a macroscopic capacitor, determined by an $RC$-time. 
But instead of only purely classical quantities, we obtain now 
expressions which contain quantum corrections due to the phase-coherent
electron motion in the cavity. It turns out 
that the $RC$-time can be expressed with the help 
of the Wigner-Smith time-delay matrix 
\begin{equation}
{\cal N} = \frac{1}{2\pi i} {\bf s}^{\dagger} \frac{d{\bf s}}{dE} . 
\label{ws}
\end{equation}
The sum of the diagonal elements of this  matrix 
determines the density of states 
\begin{equation}
N = Tr{\cal N} = \frac{1}{2\pi i} Tr[{\bf s}^{\dagger} \frac{d{\bf s}}{dE}]
\label{den}
\end{equation}
and gives rise to a "quantum capacitance" $e^{2} N$
which in series with the geometrical capacitance 
determines the electrochemical capacitance 
\begin{equation}
C_{\mu}^{-1} = C^{-1} + (e^{2} N)^{-1} . 
\label{cmu}
\end{equation}
The resistance which counts is the charge relaxation resistance (index q)
\begin{equation}
R_{q} = \frac{e^{2}}{2h} \frac{Tr[{\cal N}^{\dagger} {\cal N}]}
{[Tr{\cal N}]^{2}} .
\label{rq}
\end{equation} 
For simplicity we have given these results, Eqs. (\ref{ws} - \ref{rq})
only in the zero temperature limit. 
It is instructive to consider a basis in which the 
scattering matrix is diagonal. 
Since we have only reflection all eigenvalues of the scattering matrix 
are of the form $exp(i\phi_{n})$ where 
$\phi_{n}$ is the phase which a carrier 
accumulates from the entrance to the cavity through multiple 
scattering inside the cavity until it finally exits the cavity.
Thus the density of states can also be expressed as 
\begin{equation}
N = (1/2\pi) \sum_{n}(d\phi_{n}/dE)
\label{denp}
\end{equation} 
and is seen to be proportional to the total Wigner time delay carriers 
experience in the cavity. The time delay for channel n is 
$\tau_n = \hbar d\phi_n/dE$. 
Similarly we can express the charge relaxation resistance in terms 
of the energy derivatives of phases
and we obtain in the zero-temperature limit, 
\begin{equation}
R_{q} = \frac{e^{2}}{2h} \frac{\sum_{n} (d\phi_n/dE)^{2}}{[\sum_n d\phi_{n}/dE]^{2}}.
\label{rqp}
\end{equation} 
$R_q$ is thus determined by the sum of the squares of the delay times
divided by the square of the sum of the delay times. 
We now briefly discuss these results. 
First, our Eq. (\ref{cmu}) for the electrochemical capacitance predicts 
that it is not a purely geometrical quantity but that 
it depends on the density of states of the cavity. 
This effect is well known from investigations of the capacitance of 
the quantized Hall effect. More recent work investigates 
the mesoscopic capacitance of quantum dots and wires
and is often termed {\em capacitance spectroscopy}. In addition, 
to the average behavior our results can also be used 
to investigate the fluctuations in the capacitance. 
Similar to the universal conductance fluctuations 
there are capacitance fluctuations in mesoscopic samples
due to the fluctuation of the density of states. 
Such effects can be expected to be most pronounced 
if the contact permits just the transmission of a single channel. 
Then it is necessary not only to investigate the fluctuations 
of the mean square fluctuations but the entire distribution function. 
Such an investigation was carried out by Gopar et al. \cite{gopar}
in the single channel limit and by Brouwer and the author \cite{brbu}, 
and Brouwer et al. \cite{bfb} for chaotic cavities with quantum point contacts 
which are wide open (many channel limit). 
Since the Coulomb energy $e^{2}/C$ is typically much larger than 
the level separation $\Delta$ these fluctuations are small and 
possibly hard to observe. 

Next let us discuss briefly the charge relaxation resistance $R_q$. 
First we note that the resistance unit is not the von Klitzing 
$h/e^{2}$ but $h/2e^{2}$. The factor two arises since the cavity 
is coupled to one reservoir only. Thus only half the energy is dissipated 
as compared to dc-transport through a two terminal conductor. 
Second, we note that in the single channel limit, Eq. (\ref{rqp}) 
is {\em universal} and given just by $h/2e^{2}$. This is astonishing since 
if we imagine that a barrier is inserted into the lead connecting 
the cavity to the reservoir one would expect a charge relaxation resistance 
that increases as the transparency of the barrier is lowered. 
In the large channel limit, Eq. (\ref{rqp})
is proportional to $1/N$, where $N$ is 
the number of scattering channels, 
and it can be shown that its ensemble averaged 
value is indeed proportional to $1/T$,  
if each channel is connected with transmission 
probability $T$ to the reservoir \cite{bunp}. Thus in the large 
channel limit Eq. (\ref{rqp}) behaves as expected. 

Using the fluctuation dissipation theorem we also obtain the 
fluctuations of the current, the charge on the cavity, and the 
potential. Ref. [\citen{btp}] gives a direct derivation of these
fluctuation spectra without invoking the fluctuation dissipation theorem. 

The above discussion emphasizes the role of interaction 
in the investigation of time-dependent problems. 
The discussion also highlights that capacitances and charge relaxation
resistance, or taken together, the RC-time, are fundamental for our 
understanding of ac-transport. Clearly, the simple view taken here
which neglects exchange correlations and for poor contacts neglects 
effects which arise from the discreteness of the charge, leaves much room for 
improvement. For a discussion of capacitance fluctuations, taking 
into account the discreteness of charge we refer the reader to Kaminski
et al. \cite{kami} which builds on earlier work by Flensberg \cite{flen2} and 
Matveev \cite{matv}.

\section{The dynamic conductance matrix} 

Consider now an arbitrary geometry consisting of a mesoscopic 
conductor with $M$ contacts, $\alpha =1, 2, 3,..,M$
and $N-M$ gates, $\alpha = M+1,..,N$. We can use a similar approach 
as out-lined above \cite{bpt} or in fact an approach which 
uses the full potential landscape \cite{jpc,math} 
to determine the dynamic conductance matrix 
$G_{\alpha \beta}(\omega) = dI_{\alpha} (\omega) /dV_{\beta} (\omega)$.
For simplicity, we assume that the external circuit connecting 
the various contacts exhibits zero impedance in all branches. 
An expansion of the dynamic conductance 
to second order in frequency is,  
\begin{equation}
G_{\alpha \beta}(\omega) = G_{\alpha\beta}(0) - i \omega 
E_{\alpha\beta} + \omega^{2}  K_{\alpha\beta}
+ O(\omega^{3}) .
\label{condmat}
\end{equation} 
Here the first term is the dc-conductance. 
This matrix has non-vanishing elements only for $\alpha < M$ and 
$\beta < M$, i. e. between contacts that permit carrier transmission. 
The second term is called the {\it emittance} matrix. 
If either $\alpha > M$ or $\beta > M$ and if both $\alpha > M$ and $\beta > M$
the elements of this matrix are purely capacitive. They are 
determined by the conductor to gate capacitances and by gate-gate 
capacitances. Even if $\alpha < M$ or $\beta < M$ the elements of 
this matrix might have the same sign as expected from a capacitance matrix.
But for ballistic structures, or other structures with high transmission,
the coefficients of this matrix might be dominated by kinetic effects 
and have a sign that we would expect if the system also contains 
inductive elements. Typically only coupling to the Poisson equation is 
considered and not to the full Maxwell equations. In this case 
we call a coefficient of the emittance matrix with a sign 
opposite to what is expected for a capacitance, {\em kinetic-inductive}. 
Below, we consider an example of such a matrix. 

The term second order in frequency is dissipative and of the 
type $C^{2}_{\mu}R_q$, but again with a sign that depends 
on the kinetics of the transport. 

\subsection{The emittance matrix of a quantum point contact} 

\begin{figure}[t]
\hspace{1mm}
 \epsfxsize = 50 mm
 \epsffile{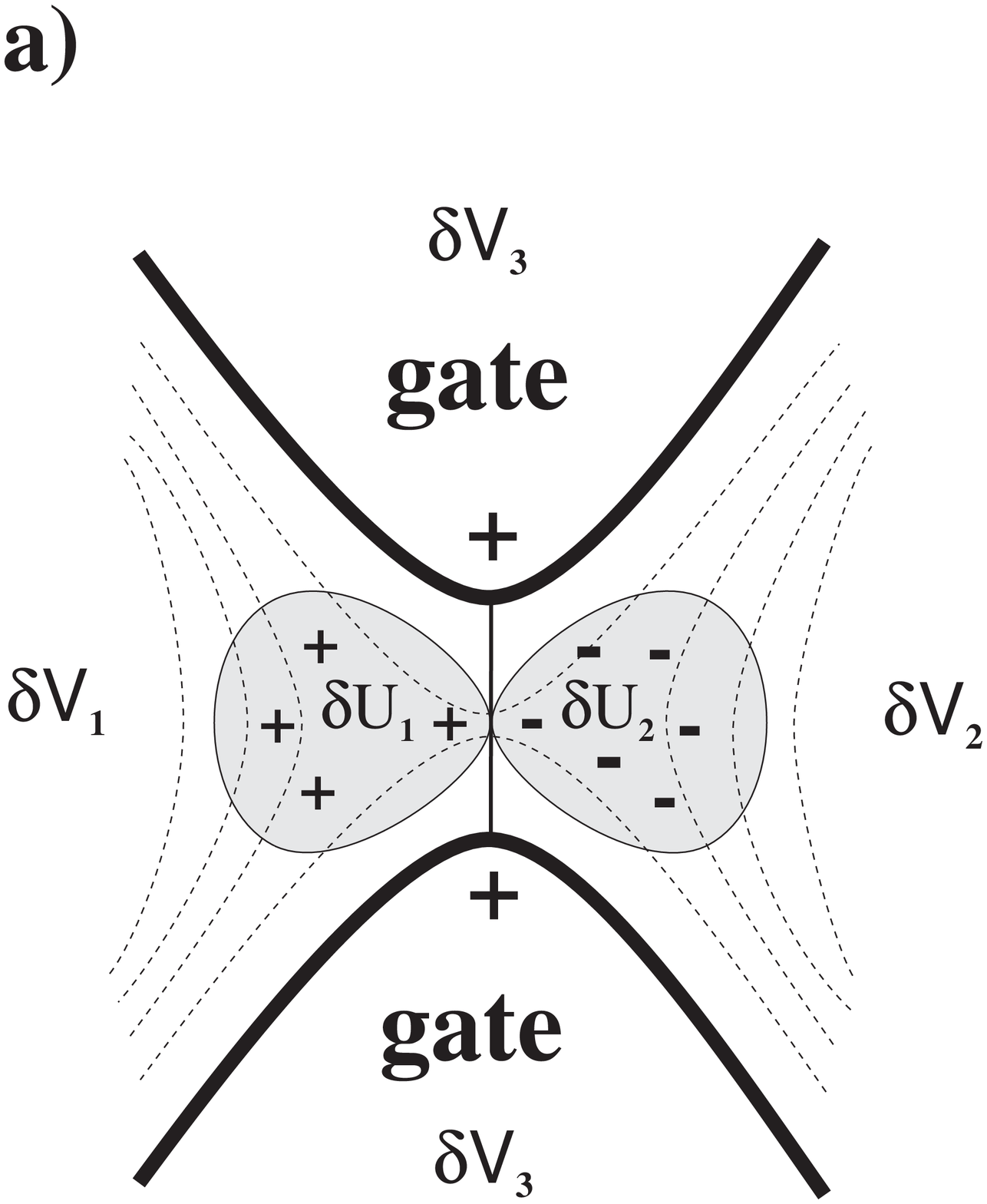}
\hspace{5mm}
 \epsfxsize = 55 mm
 \epsffile{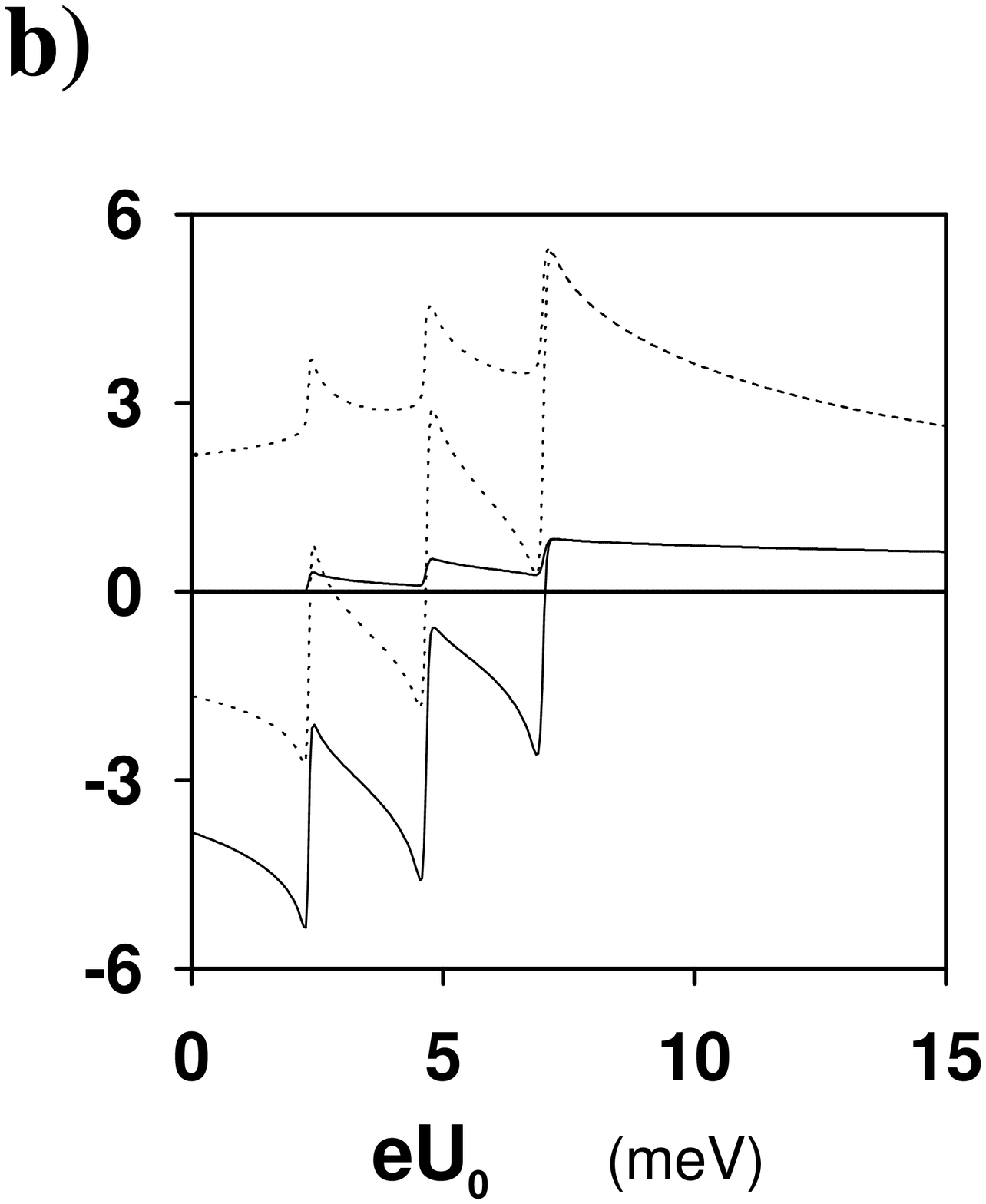}
 \vspace{-0.5cm}
\caption{ a) Quantum point contact with gates. The regions $\Omega _{1,2}$
at the left and right of the point contact can be charged equally 
vis-a-vis the gates or can be charged unequally to 
form a dipole. b) Capacitance $C_{\mu }$
and emittance $E_{11}$ of the QPC. Capacitances and emittances are in units of 
femto farads.}
\label{fig5}
\end{figure}

As an example, we consider here the emittance matrix 
of a quantum point contact (QPC). A QPC is a small constriction 
in a two-dimensional 
electron gas which allows the transmission of only a few conducting 
channels. 
We consider a symmetric QPC with two gates as shown in Fig. \ref{fig5}a. 
and ask for its
capacitance and low-frequency admittance. 
Again we greatly simplify the electrostatic 
problem by assuming that there are only two regions 
$\Omega _{1}$ and $\Omega _{2}$ 
to the left and to the right of the constriction
with sizes of the order of the screening length (see Fig. \ref{fig5}a.).
We are interested in the charge variation in these two regions. 
The theory now deals with two potentials 
$\delta U_{1}$ and $\delta U_{2}$ which describe the departure
away from equilibrium of the electrostatic potentials in these regions.  
We only present the result
that describes the opening of the first 
quantum channel. Thus the QPC has a transmission probability $T$ 
and a reflection
probability $R$. Furthermore, we assume that at equilibrium 
the QPC has a right-left symmetry. 
The two gates are taken to be at the same voltage $V_{3}$.
Thus in effect, the two gates act like a single gate.
As in the previous section, the gate will be treated as a 
macroscopic conductor. 
Charge conservation is taken into account by requiring 
that the sum of the charges
in $\Omega_{1}$, $\Omega_{2}$ and at the gates vanishes,
$dq_{1}+ dq_{2}+ dq_{3} = 0$.

The geometrical capacitance is now also a matrix $dq_{i} = C_{ij} dU_{j}$
and due to the symmetry of the problem,
the geometric capacitance matrix can be written in the form
\begin{equation}
{\bf C} = 
\left(
\begin{array}{ccc}
C_{0}+ C& -C_{0} & -C \\
-C_{0} & C_{0}+C & -C\\
-C & -C& 2C 
\end{array}
\right) 
\;\;\;,
\label{cgeo}
\end{equation}
where $C_{0}$ is the geometric capacitance between $\Omega_{1}$ 
and $\Omega_{2}$ and where $C$ is the geometric capacitance
between these regions and the gates.
The emittance matrix can be expressed with the 
help of two electrochemical capacitances $C_{g}$ and $C_{\mu }$.
Here $C_{g}$ is the electrochemical capacitance of the QPC
vis-a-vis the gate and 
$C_{\mu }$ is the capacitance which determines to which extend 
we can charge the regions $\Omega_{1}$ and $\Omega_{2}$ differently, 
i. e. build up a {\em dipole} across the QPC. Denoting the quantum contribution 
to the capacitance of the two regions by $D = e^{2} N_{\Omega_{1}+\Omega_{2}}$ 
Ref. [\citen{mbtc}] finds  
\begin{eqnarray}
C_{g} & = & \frac{1}{C^{-1}+ (D/2)^{-1} } , 
\label{Cgqpc} \\
C_{\mu} & = & \frac{ RC_{0}+ (C/2) (1+R) + 2C_{0}C_{g}D^{-1} }{1+
2(2C_{0}+ C)D^{-1}} \;\; .
\label{Cmqpc}
\end{eqnarray}
Note that $C_{g}$ is of the same form as Eq. (\ref{cmu}), 
whereas the $C_{\mu}$ now depends on the reflection probability 
of the QPC. For the emittance matrix Ref. [\citen{mbtc}] finds 
\begin{eqnarray}
E_{11} & = & E_{22} = R C_{\mu} -DT^{2}/4 + C_{g} T/2 , \nonumber \\
E_{12} & = &  E_{21} = C_{g}  - E_{11} , \nonumber \\
E_{13} & = &  E_{31} = E_{23}  =   E_{32} = - E_{33}/2 = - C_{g} .
\label{E11}
\end{eqnarray}
To elucidate the content of these equations consider the limiting case in 
which $C_{0}$ tends to zero. 
In this limit both the charge of the QPC and on the gate are 
fixed. We have $C_g = 0$ and all elements of the emittance 
matrix vanish except four elements which are equal in 
magnitude $E_{11} = E_{22} = - E_{21} = - E_{21} \equiv E$
with  \cite{tcmb} $E = R C_{\mu} -DT^{2}/4$. 
For a small transmission probability 
the first term in $E$ dominates: We have a very weakly {\em leaking capacitor}. 
On the other hand as the first channel becomes transparent and $T$ 
tends to one, we have a ballistic conductor. The emittance is 
negative and has the sign characteristic not of a capacitive response 
but of a (kinetic) inductive response. The full curve in 
Fig. \ref{fig5}b, which stays positive, is the capacitance $C_{\mu}$
as a function of the value 
$eU_{0}$ at the saddle point of the QPC potential 
and the curve which departs from this line and reaches negative values 
is $E$. The potential range shown covers the opening 
of three successive quantum channels which are separated by $E_{F}/3 = 7/3 meV$. 
The dashed lines in Fig. \ref{fig5}b. show the behavior of the 
capacitance $C_{\mu}$ and the emittance element $E_{11}$ for 
$C = C_{0}$. The features in the capacitance become smaller and the 
transition from capacitive to kinetic inductive behavior extends 
over a voltage region corresponding to the opening of several channels. 

An alternative discussion of the admittance of a QPC is presented 
by Aronov et al. \cite{aron}, 
who attempt to find the entire potential landscape. 
Strangely, however, they find regions in which the electric field 
points against the overall voltage drop. Of course that is not forbidden 
by any general principle, but for a QPC we expect the potential 
landscape to be a smoothly varying function both in the equilibrium 
state and in the presence of slowly oscillating external potentials.

\subsection{Negative capacitance?} 

Before continuing the discussion, it is worthwhile to discuss briefly
the rather entrenched practice to speak about {\em negative capacitance}
of conductors which exhibit a (kinetic) inductive response $E_{11} < 0$
rather then as expected a capacitive response $E_{11} > 0$. 
As an example we cite here only two items \cite{ersh}.
We emphasize that in a dynamical conductance 
measurement it is the emittance $E$ which is measured and not 
really the capacitance $C_{\mu}$. The example we have discussed shows that 
the capacitances $C_{\mu}$ and $C_{g}$ stay positive, independently 
of whether the emittance is positive or negative. There are 
examples for which the compressibility is negative and in such a case
the term negative capacitance might be appropriate. 

\subsection{Magnetic field symmetry of dynamic conductance} 

Geometrical capacitances are independent of magnetic field. 
Through the density of states, however, the electrochemical 
capacitance becomes magnetic field dependent. As long as it 
is only the total (global or local) density of states which counts, 
capacitance coefficients are even functions of magnetic field. 
For a conductor with $M \ge 2$ contacts, the emittance matrix 
of Eq. (\ref{condmat}) obeys the Onsager reciprocity symmetry 
$E_{\alpha\beta} (B) = E_{\beta\alpha} (-B)$. It can be shown 
that this symmetry relation applies also to the purely 
capacitive elements in the emittance matrix and 
that therefore capacitance elements exist which are not 
even functions of magnetic field \cite{mb93}.  
Experiments demonstrating this for a geometry where a 
small gate overlaps the edge of a two-dimensional conductor 
with two contacts have been carried out by Chen et al. \cite{chen}
in the integer regime and by Moon et al. \cite{moon} in the fractional
quantized Hall regime. In contrast, if we consider an arrangement of 
conductors each of which is connected by only {\em one} lead to a reservoir, 
the emittance matrix (which in this case is a pure capacitance 
matrix) is an even function of magnetic field. Under the same 
condition $K_{\alpha\beta}$ is also an even function of magnetic 
field but only as long as inelastic scattering can be neglected.   
This later point, the change of symmetry depending on inelastic
scattering is clearly intersting and deserves further work. 
An experiment on such a geometry is reported in Ref. [\citen{somm}].
A classification of the magnetic field symmetry is given in Ref. [\citen{chri}].

\subsection{Frequency-dependent weak localization} 

In a pioneering experiment Pieper and Price \cite{piep}
investigated the dynamic conductance of a mesoscopic 
Aharonov-Bohm 
ring with frequencies up to the GHz range. 
Denoting the time for diffusion across 
the metallic ring by $\tau_{d} = L^{2}/D$, the frequency is large enough 
to measure the real and imaginary 
part for $\omega \tau_{d} < 1$ 
and $\omega \tau_{d} >  1$. A drawback of the experiment is 
that the temperature $kT > \hbar \omega$ at all accesible frequencies. 
As a consequence, as discussed in Refs. [\citen{pie2,staf}], 
the amplitude of the Aharonov-Bohm oscillations is nearly frequency 
independent. Below, I discuss briefly the frequency dependence of the 
weak localization in quantum chaotic cavities discussed in the charge 
neutral limit by Aleiner and Larkin \cite{alla} and for a chaotic cavity 
in proximity to a gate by Brouwer \cite{brbu} and the author. 

Ref. [\citen{brbu}] proceeds by first evaluating the 
conductance matrix, Eq. (\ref{0con}), in the absence of screening. 
The currents are evaluated in response to an oscillating voltage
in the contacts under the condition that the potential in the cavity 
is held fixed. 
For a quantum chaotic cavity coupled to reservoirs
via two large contacts with $N_{1}$ and $N_{2}$ channels,
and using random matrix theory [\citen{bb96a}]
to perform the ensemble averages, 
Ref. [\citen{brbu}] finds in an expansion up to order $1$ in $N^{-1}$, 
where $N = N_{1} +N_{2}$, 
for the ensemble averaged conductance 
$\langle G_{\mu\nu}^{u}(\omega) \rangle $ 
the following results. To leading order (order  $N$) there is a 
classical contribution 
\begin{eqnarray} \label{eq:Tabavg}
  \langle G_{\mu\nu}^{cl,u}(\omega) \rangle = \delta_{\mu \nu} N_{\mu} - 
  {N_{\mu} N_{\nu} \over N(1 - i \omega \tau_{d})}, 
\end{eqnarray}
and to order $1$ there is a weak localization correction 
\begin{eqnarray} \label{eq:Tabwl}
\langle \Delta G_{\mu\nu}^{u}(\omega) \rangle = {(2-\beta)N_{\mu} 
  \over \beta N (1 - i \omega \tau_{d})} 
  \left({N_{\nu} 
  (1 - 2 i \omega \tau_{d}) \over N (1 - i \omega \tau_{d})^2} - 
  \delta_{\mu\nu} \right),
\end{eqnarray}  
where $\tau_{d} = (h/N) \langle dn/d\varepsilon \rangle =(h/N \Delta) $ 
is the dwell time
and $\Delta$
is the mean level spacing. The index $u$ indicates that we deal with an unscreened 
conductance. The symmetry index $\beta=1$ ($2$) in the absence (presence) of a time-reversal-
symmetry breaking magnetic field; $\beta=4$ in zero magnetic field with strong 
spin-orbit scattering. The matrix  
$\langle G_{\mu\nu}^{u}(\omega) \rangle  = 
\langle G_{\mu\nu}^{cl,u}(\omega) \rangle  
+ \langle \Delta G_{\mu\nu}^{u}(\omega) \rangle$ 
is not current conserving. 
Next let us compare this result with the case 
when screening is taken into account. Coupling to a nearby gate is again 
described by a geometrical capacitance. 
Ref. [\citen{brbu}] finds for the screened admittance 
$\langle G_{\mu\nu}(\omega) \rangle$,
the classical contribution 
\begin{eqnarray} \label{eq:Gadmavgcl}
  \langle G^{cl}_{\mu\nu}(\omega) \rangle = 
  \delta_{\mu \nu} N_{\mu} - {N_{\mu} N_{\nu} 
  \over N(1 - i \omega \tau)},
\end{eqnarray} 
and the weak localization contribution 
\begin{eqnarray} \label{eq:Gadmavgwl}
  \langle \Delta G_{\mu\nu}(\omega) \rangle = 
 {(2-\beta) N_{\mu} 
  \over \beta N (1 - i \omega \tau_{d})}
  \left({N_{\nu} (1 - 2 i \omega \tau) \over 
  N (1 - i \omega \tau)^2} - \delta_{\mu\nu} \right),
\end{eqnarray} 
where $\tau^{-1} = \tau_{d}^{-1} + e^2 N/h C$ is the 
$R_qC_{\mu}$ time. 
The electrochemical capacitance of the cavity is 
$C^{-1}_{\mu} = C^{-1} + (e^{2} \langle dn/d\varepsilon \rangle )^{-1} =
C^{-1} + (N e^{2} \tau_d/h)^{-1}$ 
and the charge relaxation resistance is
\begin{eqnarray}
  R_{q} = \frac{h}{e^{2}} \frac{1}{N} =  
  \frac{h}{e^{2}} \frac{1}{N_{1}+N_{2}} 
\label{eq:rqcav}
\end{eqnarray} 
For the product we have $R_qC_{\mu} = \tau$. 
It is intersting to compare the charge relaxation resistance $R_q$ 
with the dc-resistance $R = ({h}/{e^{2}}) (1/N_{1} +1/N_{2})$
which is the {\em series} addition of the contact resistances and 
is thus dominated by the smaller of the two contacts. 
In contrast, the inverse of $R_q$
is the {\em parallel} addition of the contact conductances, and 
$R_q$ is thus dominated by the larger of the two contacts.

Comparison of the two results shows that screening leads almost everywhere 
to the replacement of the dwell time $\tau_d$ by the $RC$-time $\tau$. 
The dwell term survives only in the weak localization correction 
which depends on both time-scales. That the dwell time can survive 
in the weak localization term is explained by 
the fact that the time reversed paths which give rise to weak localization 
can be viewed as (charge neutral) electron-hole trajectories. 
Since in typical experiments the charging energy is a few times larger 
than the level spacing the weak localization term depends on vastly 
different time scales $\tau << \tau_d$ and this double time-scale 
behavior should be observable in experiment. 

The conductance matrix, Eqs. (\ref{eq:Gadmavgcl}, \ref{eq:Gadmavgwl}) 
is current conserving if the gate contact is included. 
The elements of the conductance matrix relating to the gate 
(contact 3) can be obtained by using the sum rules 
$\sum_{\mu} G_{\mu\nu} = \sum_{\nu} G_{\mu\nu} = 0$. 
Of course we can also insist that the unscreened 
result is current conserving and apply the above sum rules 
to determine the remaining elements of the conductance matrix. 
That corresponds to a cavity which has an infinite capacitance 
towards the gate.

\subsection{The pulsed cavity}

Consider now an experiment where we apply a voltage pulse 
$V_{\alpha} (t) = a_{\alpha} \delta(t)$
to one of the contacts. The Fourier transform of 
such a pulse is
a constant and the resulting current 
is thus proportional to the frequency integral of the conductance 
matrix element. The current at contact $\mu$ in response to a pulse 
at contact $\nu$ is thus proportional to the frequency integral 
over the conductance $G_{\mu\nu} (\omega)$, or 
$G_{\mu\nu}(t) 
= (1/2\pi) \int d\omega exp(-i\omega t) G_{\mu\nu}(\omega)$. 
The aim of the discussion presented here is two fold. 
First, we would like to establish the connection 
between Ref. [\citen{brbu}] and works \cite{numb} which discuss
a novel time scale which is intermediate between the dwell time 
$\tau_d$ and the Heisenberg time $\tau_H$. The Heisenberg time 
is  $\tau_H = N \tau_d$. It is the time scale at which a quantum system
starts to notice the energy level structure. This connection 
concerns only the non-interacting system. Our second aim is to point 
out that the interacting system behaves differently and the 
intermediate time mentioned above does not appear, at least not within 
the limits of the first two terms in the $1/N$-expansion of the 
conductance. 

For the classical part of the ensemble averaged conductance 
this gives for the non-interacting system (for $\mu =1,2, \nu =1,2$)
\begin{eqnarray}
  \langle G^{cl,u}_{\mu\nu}(t) \rangle  = \frac{N_{\mu}N_{\nu}}{N\tau_d}
  exp(-t/\tau_d)
\end{eqnarray}
an exponential decay determined by the dwell time, 
whereas for the interacting system 
the decay is very much faster since it is determined by $\tau$,
\begin{eqnarray}
  \langle G^{cl}_{\mu\nu}(t) \rangle  = \frac{N_{\mu}N_{\nu}}{N\tau}
  exp(-t/\tau). 
\end{eqnarray}
On a log scale we have 
$ln [\langle G^{cl,u}_{\mu\nu}(t) \rangle N/N_{\mu}N_{\nu}] = -t/\tau_d$
for the non-interacting system and 
$ln [\langle G^{cl}_{\mu\nu}(t) \rangle N/N_{\mu}N_{\nu}] = -t/\tau$
for the classical, interacting system. 

Let us next investigate the weak localization 
term. For the non-interacting case (index u), we find 
\begin{eqnarray}
ln [\langle \Delta G^{u}_{\mu\nu}(t) \rangle  \frac{N}{N_{\mu}N_{\nu}}] = -t/\tau_{d}
+ ln [\frac{2-\beta}{N_{\nu}\beta} (\delta_{\mu\nu} - 
\frac{N_{\nu}}{N}(\frac{t}{\tau_d})(2- \frac{t}{2\tau_d}))] .
\end{eqnarray}
The time-dependence is now already somewhat complicated as a consequence 
of the fact that the pole determined by the dwell time is of third order 
in the weak localization term (see Eq. (\ref{eq:Gadmavgwl})). 
While the initial time-dependence 
is governed by $\tau_d$ there exists for the orthogonal and symplectic
examples, a regime when the last term in the parenthesis proportional 
to $t^{2}$ becomes dominant and a deviation from simple exponential 
behavior becomes observable. The term proportional to $t^{2}$ 
will be of order 1 at a time $t^{2} = t^{2}_g =N \tau_{d}^{2} = \tau_d \tau_H$.
where $\tau_H = N \tau_d$ is the Heisenberg time. 
This establishes the connection between the results of Ref. [\citen{brbu}] 
and the discussions in Refs. [\citen{numb}].  

For the interacting system, there is now an additional 
time scale $\tau$ in the weak localization term in addition to the dwell 
time $\tau_d$. As a consequence, the pole in the weak localization 
term determined by the dwell time is only first order. The result is 
that even if screening is not complete (finite capacitance)
there are no prefactors which grow proportional to $t^{2}$, but 
only with $t$. The result is simple only in the limit of complete 
screening ($\tau = 0$), where the decay of the weak localization term is 
exponential with the dwell time. We have 
$G_{11}(t) = G_{22}(t) = 
- G_{12}(t) = - G_{21}(t)  \equiv G(t)$ 
with 
\begin{eqnarray}
  \langle G(t) \rangle &=& G^{dc} 
  [\delta(t) + \frac{2-\beta}{\beta} \frac{1}{N} \frac{1}{\tau_d}
  exp(-t/\tau_d)] 
\end{eqnarray} 
where $G^{dc}$ is the dc-conductance of the cavity. 
Note that in this case the classical response is instantaneous
(within our approximations).

\section{Pumping}

\subsection{Adiabatic quantum pumping}

Different types of electron pumps have been of interest for 
a number of years. 
Adiabatic quantum pumping investigates the 
current in response to two (slowly) oscillating potentials 
$U_{1} (t) = u_{1} \, 
sin(\omega t)$ and 
$U_{2} = u_{2} \, sin(\omega t - \phi)$ which in
practice are applied to the system by varying two 
gate voltages. The pumping is called adiabatic since the frequencies 
are sufficiently slow for the system to follow a quasi-stationary state. 
It is called quantum pumping since the direction and magnitude of the current
that is pumped depend on the sample specific quantum nature of the 
electron wave functions. 
Different theoretical 
approaches have been put forth for metallic diffusive conductors \cite{spiv}  
and chaotic cavities with perfect contacts \cite{brpu} and with 
contacts which 
are almost transparent \cite{al2}. 
Below I present some results of the work of 
Brouwer \cite{brpu}. The experiment by Switkes et al. \cite{swit} 
investigates the current response
as a function of the phase difference $\phi$. 

The charge which is expelled through contact $\alpha$ 
in response to an oscillating potential $U_{1}$ is given by 
\begin{equation}
\delta Q_{\alpha} (t) = e^{2} N(\alpha,1) \delta U_{1}(t) 
\label{exp}
\end{equation}
where  
\begin{equation}
N(\alpha,1) = - (1/4\pi  i e) \sum_{\beta}
Tr({\bf s}^{\dagger}_{\alpha\beta} (\delta{\bf s}_{\alpha\beta}/\delta U_{1}) - 
(\delta{\bf s}^{\dagger}_{\alpha\beta}/\delta U_{1} )
{\bf s}_{\alpha\beta} )
\end{equation}
is the emittance \cite{btpz,mb93} of the conductor into contact $\alpha$. 
For the case of interest here, in the presence of two potentials 
the charge emitted through contact $\alpha$ is 
\begin{equation}
\delta Q_{\alpha} (t) = e^{2} (N(\alpha,1) \delta U_{1}(t) +
N(\alpha,2) \delta U_{2}(t)) .
\label{exp1}
\end{equation}
The total charge expelled in a time interval from $0$ to $T$ is 
\begin{equation}
dQ_{\alpha} (t) = e^{2} \int_{0}^{T} dt (N(\alpha,1) dU_{1}/dt +
N(\alpha,2) dU_{2}(t)/dt) .
\label{exp2}
\end{equation}
For the case considered here, where $U_{1}$ and $U_{2}$ are periodic
functions in time with period $T$, 
the pair describes in this parameter space a closed path 
$S$.  With the help 
of Green's theorem, this integral can be written as a surface integral
of the surface enclosed by the path $S$, 
\begin{equation}
dQ_{\alpha} (t) = e^{2} \int_{S} 
(\frac{\partial} {\partial U_{1}} N(\alpha,2) -
\frac{\partial} {\partial U_{2}} N(\alpha,1)) dU_{1}dU_{2} .
\label{exp3}
\end{equation}
Explicitly, in terms of the scattering matrices this result becomes
\begin{equation}
Q_{\alpha} (T) = - (\frac{e}{2\pi  i}) \sum_{\beta}\int_{S} 
Tr[\frac{\delta{\bf s}^{\dagger}_{\alpha\beta}}{\delta U_{1}} 
\frac{\delta{\bf s}_{\alpha\beta}}{\delta U_{2}}- 
\frac{\delta{\bf s}^{\dagger}_{\alpha\beta}}{\delta U_{2}} 
\frac{\delta{\bf s}_{\alpha\beta}}{\delta U_{2}}]dU_{1}dU_{2} . 
\label{exp4}
\end{equation}
The resulting current at contact $\alpha$ is 
\begin{equation}
I_{\alpha}  = \frac{e \omega \, sin(\phi) u_{1} u_{2}}{2 \pi} 
\sum_{\beta}
Im Tr[(\delta{\bf s}^{\dagger}_{\alpha\beta}/\delta U_{1}) 
(\delta{\bf s}_{\alpha\beta}/\delta U_{2} )]. 
\label{exp5}
\end{equation}
Experimentally, the potentials $U_{1}$ and $U_{2}$ are not known.
What is controlled are the voltages applied to the gates. 
Thus the considerations above should be extended in this direction. 
Brouwer \cite{brpu} does consider screening and finds for the problems
he investigates only hardly noticeable changes compared to the unscreened result. 

An experiment testing these predictions was performed by 
Switkes et al. \cite{swit} for a quantum chaotic cavity
which is connected to reservoirs via 
quantum point contacts which are completely transparent for the lowest 
quantum channel. Oscillating 
voltages are applied to two of the gates used to define the geometry of 
the cavity.
Switkes et al. measure the voltage which is produced by the pumping 
in an infinite external impedance circuit. The voltage is
described by the simple expression $V_{dot} = A_{0} sin(\phi) +B_{0}$
with $A_{0}$ and $B_{0}$ extracted from fits to the data. 
It is found that $A_{0}$ fluctuates randomly (as a function of magnetic field)
with an average that is roughly forty times smaller 
than the fluctuation amplitude. Similarly $B_{0}$ is a very small correction. 
For low pumping amplitudes the experimental data agree well with theory. 
We conclude the description of this adiabatic pump by emphasizing its 
wide applicability to test phase coherent fluctuation properties in a wide 
range of mesoscopic systems. 

\subsection{Mechanical pumping}

Thus far we have only considered electrical degrees of freedom. 
It is quite interesting that during recent years, different pumping 
mechanisms have been investigated, which are based on mechanical 
pumping. The pumps based on launching surface acoustic waves 
through a mesoscopic sample provide one example. Another example
are shuttle pumps investigated by Gorelik et al. \cite{gor} : 
a quantum dot oscillates between 
the left and right 
reservoir and during each oscillation transfers one charge. 
Experiments are reported 
in Ref. [\citen{erbe}]. The accuracy of such a pump 
is the subject of work by Weiss and Zwerger \cite{zwe}. 
Such developments are important, especially in view of 
current and capacitance standards \cite{flen}. 
They are scientifically interesting since the 
domain of electro-mechanical
mesoscopic effects remains largely unexplored. 

\section{Photon-assisted transport}

\subsection{Quantum dots} 

A successful set of experiments has been carried out 
by applying microwaves in the GHz range to quantum dots. 
Whereas earlier experiments dealt with quantum dots for which the 
density of states could effectively be viewed as continuous, recently
an experiment by Oosterkamp et al. \cite{oos97} succeeded in effect to 
see photon-assisted transport via the ground state and or through an 
excited state. Photon-assisted transport is a non-linear 
mechanism whereby a carrier absorbs and re-emits a photon and 
a dc-current is generated. In the dots investigated in Ref. [\citen{oos97}]
the charging plays an important role. Theoretically such a situation was 
treated by Bruder and Schoeller \cite{brsc} extending a master equation approach 
to the Coulomb blockade into the dynamic regime. 
The experimenters compare their results with the discussion 
of Tien and Gordon of photon-assisted transport (PAT) which 
states that the essential effect of the ac-voltage drop 
over the tunneling barrier (forming the contact of the dot to the reservoir)
is to modify the static tunneling rate $\Gamma (E)$ according to 
\begin{equation}
\Gamma_{PAT} (E) = \sum_{n} J_{n}^{2}(\alpha)\Gamma (E+n \hbar \omega)
\label{pat1}
\end{equation}
where $J_{n}$ is the n-th order Bessel function and $\alpha =eV/\hbar \omega$
with $V$ the voltage drop accross the barrier. Note that the Tien and Gordon 
approach 
predicts that the current through a barrier in the presence 
of a microwave field can be obtained from the static $I-V$-characteristic. 
In Ref. [\citen{oos99}], Eq. (\ref{pat1}) is used to formulate 
a master equation for a two state charge model (corresponding to a
dot with $N$ and $N+1$ electrons). Note that for $N =2$ distributed over five 
single particle levels this gives already ten equations for the probabilities 
of all different dot configurations.  To compare with experiments 
voltage drops of different magnitude are permitted across the contacts. 
This implies that at least one reservoir potential is taken to be oscillatory 
in addition to the central potential in the dot. (We note that this is 
in contrast
to most of the theoretical work, which treats this problem by assuming 
that it is only the dot potential or the energy levels in 
the dot which oscillate). By and large the model works remarkably well. 

\subsection{Quantum Point Contacts} 

It is instructive to compare the success of 
experiments in quantum dots with similar experiments aiming at observing 
photon-assisted transport in quantum point contacts (QPC). 
Despite a number of efforts, no clear signature of photon-assisted 
transport has been observed in QPC's. Instead it seems to be possible 
to explain the experiments in terms of heating generated by the 
microwaves \cite{alm}. 
The success in quantum dots is likely due to the fact that it is 
a problem with high tunnel barriers which help to generate 
localized electric fields and help to isolate the dots from the leads \cite{yak}. 
There are many theoretical works on photon-assisted
QPC's and wires \cite{many}. It would thus be very interesting to find a way 
to observe photon-assisted transport in such structures
to test these predictions. A recent work
suggests that one should try to localize the electric field for instance 
with the help of superconductors placed above the QPC with only a narrow 
opening \cite{tach}. We remark here only, that it is not sufficient to localize 
the external field: what counts is the {\em total} field. In a QPC or a 
quantum wire the poor screening will generate a total field over a large 
region even if the external field is well localized. 

\subsection{Role of displacement currents}

In a number of works it is stated that displacement currents play no role 
in photon-assisted transport. This is correct in the sense that if 
we evaluate a dc-current the displacement part of the current 
$(\epsilon_{L}/4\pi) 
\partial {\bf E}/\partial t$ does of course not contribute. 
However, the particle current $j_{p}$ is a function of the total 
electric field and moreover, it is a non-linear function of the field. 
To investigate the role of the displacement current in photon-assisted 
transport, it is useful to remember, that even so we are mainly 
interested in the dc-current, photon-assisted transport is 
also associated with time-dependent currents. In the presence 
of sinusoidal applied voltages $dV_{\alpha}(t) = dV_{\alpha} (\omega)
cos(\omega t)$, the resulting 
currents at contact $\alpha$ have components at all harmonics of $\omega$. 
In other words a theory is needed not only for $I_{\alpha} (\omega = 0)$
but also for $I_{\alpha} (n\omega)$. The Fourier comonents $n= \pm 1$ are 
the currents at the driving frequency. Ref. [\citen{ped}] investigated this
and presents a theory based on an RPA screening approach, 
in terms of an expansion of the currents 
in powers of the Fourier amplitude of the driving voltages. 
This theory is formulated such that the overall charge 
of the conductor and gates obeys Eq. (\ref{qcons}) for all harmonics. 
It is shown, 
that the self-consistent potential within the dot is essentially 
determined by the currents and charges at $ n = \pm 1$. The self-consistent
potential in turn also determines the zero-frequency Fourier component.  
In fact, one can view photon-assisted transport as a down conversion of
the displacement fields at frequency $\omega$. RPA might not be 
the proper approach to treat problems in which charge quantization is 
important, but the central point made in Ref. [\citen{ped}] is clearly 
independent of the self-consistent scheme that is applied. 

\subsection{Pumping with a single localized potential?}
 
Another remark seems here appropriate: 
consider a quantum dot at zero external bias. 
Switkes et al. \cite{swit} write:
"A periodic deformation (of the potential) that depends on a single 
parameter cannot result in net transport; any charge that flows during 
the first half-period will flow back during the second". Parts of the 
literature are, however, in contradiction to this very plausible statement. 
The statement is obviously true, for the adiabatic quantum pump described 
above where two 
potentials are essential and in addition their mutual phase is of importance. 
It is also true for the analysis of Ref. [\citen{ped}], where a single 
spatially localized potential 
does not lead to a dc-current in the absence of a bias due to the unitarity 
of the scattering matrix. However, neither of these discussions can be 
viewed as a general proof. A proof could start along the following lines: 
potential oscillations, even in the equilibrium state, occur spontaneously,
due to thermal (or zero point quantum) noise. Such spontaneous fluctuations,
clearly, cannot lead to a dc-current since that would be tantamount to 
say that an equilibrium state does not exist. Until some 
very special conditions are fulfilled which do not correspond 
to what is possible in an equilibrium ensemble, there can be no 
resulting dc-current. 
Of course the fluctuations 
which invoke two potentials, 
as in the adiabatic quantum pump, also occur spontaneously, but with a phase 
that is random.  

\subsection{Double Dots} 

Coupling two quantum dots hybridizes the states of each dot and gives 
rise to coupled states which represent a covalent bonding and 
anit-bonding state. Two states with energies $E_l$ and $E_r$ on the left 
and the right dot in the absence of coupling are separated 
by an energy $\Delta E^{\ast} = E_{anti-bond} - E_{bond} = 
\sqrt{(\Delta E)^{2} +4|t|^{2}}$ where $\Delta E = E_l - E_r$
and $|t|$ is a coupling energy. A theoretical 
investigation of transport through double quantum dots 
is provided by Stoof and Nazarov \cite{stoo} and by 
Stafford and Wingreen \cite{sta2}. Electron transport is possible 
when an electron in the bonding state absorbs a photon and 
is promoted to the anti-bonding state. The condition for this 
process is $\hbar \omega = \Delta E^{\ast}$ or 
\begin{equation}
\Delta E = 
\sqrt{(\hbar \omega)^{2} - 4 |t|^{2}} .
\label{pat2}
\end{equation}
An experiment by Oosterkamp et al. \cite{oos3}, 
taking the coupling energy $|t|$ as a fit parameter,   
shows a remarkably good agreement with Eq. (\ref{pat2})
over a wide range of frequencies. In view of the discussion 
given above, oscillations of the 
charge in the leads, displacement currents, 
etc., it is an even more astonishing result. 
In part this can be explained by the fact that Eq. (\ref{pat2})
is a resonance condition, and that all interesting effects are thus 
buried in the coupling energy $|t|$. Clearly, it would be intersting 
to compare theoretical predictions for the width of the resonance 
with the measurements. Further, it would be intersting to see real 
time-oscillations of the excited state, as this was done 
recently in a Jospehon junction circuit by Nakamura et al. \cite{naka}. 

\section{Breadth of the field}

In the previous paragraphs we have been able to touch on a few 
problems related to dynamical transport in mesoscopic structures. 
However, a number of important topics have been omitted. 
Following is a list of a few topics
which give an impression of the wide range of questions 
addressed in this field. The references given 
are in no way complete but are presented here only as an initial guide 
to a citation trail which the interested reader has to follow on his 
own initiative. 

\subsection{Closed Systems}

Mesoscopic systems can be enclosed in the dielectric medium of 
a capacitor or a transmission line. 
This permits a contactless 
investigation of a number of susceptibilities of closed mesoscopic systems. 
We mention here in particular the work of Noat et al. \cite{noat}
where the orbital magnetic susceptibilities of closed squares 
are investigated and the work of 
Reulet et al. \cite{reul} 
where the conductance of isolated rings 
is measured in a frequency range between 330MHZ 
and 1065 MHz. Nonlinear dynamic effects in rings, such as transport 
in presence of a linearly increasing flux continue to be a subject 
of theoretical interest \cite{gro2}. 

\subsection{Aharonov-Bohm effect in capacitance}

We have pointed out that the capacitance of mesoscopic structures
is not a purely geometrical quantity but via the density of states 
depends on the properties of the system. For a small ring threaded 
by an Aharonov-Bohm flux, the charge distribution will in general 
depend on the flux, and as a consequence a capacitance can like the 
conductance of a ring exhibit Aharonov-Bohm oscillations \cite{scr}. 
Systems for which the charge quantization is important are predicted to 
exhibit especially pronounced oscillations \cite{mbcs,kriv,sima,mosk}.

\subsection{Weakly non-linear ac-response}

An intersting regime for ac-transport is the onset of
non-linearity, i. e. the initial departure away 
from Ohm's law. In this regime one might still hope 
to find answers of considerable generality. 
We refer here only to one recent theoretical work 
by Ma et al. \cite{ma}.

\subsection{Resonant Double Barriers} 

The dynamics of resonant double barrier structures has long been a subject 
of interest. We refer to Ref. [\citen{mbtc}] and the work of
Anantram \cite{anan} for a self-consistent discussion of ac-conductance.

\subsection{ac-conductance of wires} 

Perfect wires provide, like quantum point contacts and resonant double 
barriers another elementary system which can be investigated to test 
essential ideas. While the dc-conductance depends only on the equilibrium 
electrostatic potential and for an adiabatic connection to 
reservoirs is quantized, the ac-conductance is very sensitive 
to interactions. The single channel wire is very often regarded as an
example of a Luttinger liquid and addressed with the help of bosonization 
techniques \cite{safi,pono} but a discussion within RPA leads to the 
same results and is instructive \cite{bhb}. The determination of the 
coupling constants of a wire in proximity to a gate with the help 
of ac-measurements is the subject of Ref. [\citen{bhb}]. Wires connected 
to reservoirs only (no gates) but with a realistic 
Coulomb interaction are the subject of Ref. [\citen{sab}].    

\subsection{Superlattices} 

In the dynamic regime superlattices exhibit a wide variety 
of effects \cite{keay,hof}:
dynamical localization of carriers, absolute negative conductance, 
current harmonics generation, Shapiro steps, continue 
to be of interest \cite{plat}. Superlattices are 
of discussed also as THz-photon detectors \cite{igja}.

\subsection{Dynamics of edge states} 

An essential aspect of the dynamics of electrical conductors are 
plasmons. Plasmons are especially interesting in high magnetic 
fields where they propagate along the edges of the sample. 
We mention here only the experimental works by Zhitenev, et al. \cite{zhit} 
who measure the time-delay of a voltage pulse, and the work by 
Talyanskii, et al. \cite{taly} who investigate the scattering 
of edge plasmons at a barrier.

\section{Conclusion}

The investigation of dynamic transport permits us to 
probe the inner energy scales of a conductor, especially those 
associated with the charge distribution and its (collective) dynamics. 
The ratio of experiments to theoretical works and proposals is still 
very small.  It is hoped that this article can 
contribute to change this.

\section*{Acknowledgements}

This work is supported by the Swiss National Science Foundation.

\end{document}